\documentclass[epsfig,useAMS, usenatbib]{mn2e}

\usepackage{url}

\usepackage {graphics}
\usepackage{graphicx}
\usepackage {layout}

\newcommand{\Msun}{M_\odot}

\newcommand{\beq}{\begin{equation}}
\newcommand{\eeq}{\end{equation}}
\newcommand {\hi} {H\,{\small I}\,}
\newcommand {\mg} {Mg\,{\small II}\,}

\newcommand\aj{AJ}
\newcommand\apj{ApJ}
\newcommand\apjl{ApJ}
\newcommand\apjs{ApJS}
\newcommand\mnras{MNRAS}

\title[Hot gas around galaxies]{Redistributing hot gas around galaxies: do cool clouds signal a solution to the overcooling problem?}

\author[Kaufmann et al.]
{Tobias Kaufmann$^{1}$ \thanks{E-mail: tobias.kaufmann@uci.edu}, James S. Bullock$^{1}$, Ariyeh H. Maller$^2$, Taotao Fang$^{1}$\newauthor{and James Wadsley$^3$}  
\\$^1$ Center for Cosmology, Department of Physics and Astronomy, University of California, Irvine, CA 92697
\\$^2$ Dept. of Physics, New York City College of Technology, CUNY, NY, 11201 
\\$^3$ Department of Physics \& Astronomy, McMaster University, 1280 Main St.
West, Hamilton ON L8S 4M1 Canada}

\begin{document}

\pagerange{\pageref{firstpage}--\pageref{lastpage}} \pubyear{} 

\maketitle

\begin{abstract}
  We present a pair of high-resolution smoothed particle hydrodynamics
  (SPH) simulations that explore the evolution and cooling behavior of
  hot gas around Milky-Way size galaxies.  The simulations contain the
  same total baryonic mass and are identical other than their initial
  gas density distributions.  The first is initialised with a {\em low
    entropy} hot gas halo that traces the cuspy profile of the dark
  matter, and the second is initialised with a {\em high-entropy} hot
  halo with a cored density profile as might be expected in models
  with pre-heating feedback.  Galaxy formation proceeds in
  dramatically different fashion depending on the initial setup.
  While the low-entropy halo cools rapidly, primarily from the central
  region, the high-entropy halo is quasi-stable for $\sim 4$ Gyr and
  eventually cools via the fragmentation and infall of clouds from
  $\sim 100$ kpc distances. The low-entropy halo's X-ray surface
  brightness is $\sim 100$ times brighter than current limits and the
  resultant disc galaxy contains more than half of the system's
  baryons.  The high-entropy halo has an X-ray brightness that is in
  line with observations, an extended distribution of
  pressure-confined clouds reminiscent of observed populations, and a
  final disc galaxy that has half the mass and $\sim 50\%$ more
  specific angular momentum than the disc formed in the low-entropy
  simulation.  The final high-entropy system retains the majority of
  its baryons in a low-density hot halo.  The hot halo harbours a
  trace population of cool, mostly ionised, pressure-confined clouds
  that contain $\sim 10 \%$ of the halo's baryons after 10 Gyr
  of cooling.  The covering fraction for \hi and \mg absorption clouds
  in the high-entropy halo is $\sim 0.4$ and $\sim 0.6$, respectively,
  although most of the mass that fuels disc growth is ionised, and
  hence would be under counted in \hi surveys.

\end{abstract}


\section{Introduction}
\label{sec:intro}
It is well known that in the absence of feedback the majority of
baryons in galaxy-size dark matter haloes ($M \sim 10^{12}$
M$_\odot$) should have cooled into halo centres over a Hubble time
(e.g. White \& Rees 1978; Katz 1992; Benson et al. 2003).  In
contrast, only $\sim 20 \%$ of the associated baryons in Milky-Way
size haloes are observed to be in a cold, collapsed form (Maller \&
Bullock 2004 (MB04); Mo et al. 2005; Fukugita \& Peebles 2006;
Nicastro et al. 2008).  An understanding of the feedback processes
that act to solve this galaxy overcooling problem is a major goal
of galaxy formation today.  It is not known if the undiscovered galactic baryons
exist primarily in hot gaseous halos around normal galaxies (MB04; Fukugita
\& Peebles 2006; Sommer-Larsen 2006) or if they have been mostly
expelled as a result of energetic blow-out (e.g., Dekel \& Silk 1986; Oppenheimer \&
Dav{\'e} 2006).

There are several reasons to take seriously the possibility that a large fraction of the missing galactic
baryons reside in the halos of normal galaxies.
In the Milky Way, 
X-ray absorption lines produced by local hot gas are detected in the spectra of several bright AGN (e.g., Williams et al., 2005; Fang et al., 2006).  Many argue that this 
absorption corresponds to local gas ( $\sim 50$ kpc; e.g., Wang et al., 2005;  Fang et al., 2006; Bregman \& 
Lloyd-Davies, 2007), but the true origin of these features, including whether it is associated with a hot component of the Milky Way disc or with an extended hot halo, remains open to debate. Interestingly, 
Sembach et al. (2003) and Tripp et al. (2003 ) have argued that high-velocity features observed by FUSE highlight the boundaries between HI clouds (High-Velocity Clouds, HVCs) and an 
extended, hot gaseous corona around the Galaxy. Further indirect evidence for a hot Galactic 
halo comes from ram-pressure stripping models the Magellanic stream (e.g., Mastropietro et al. (2005); Kaufmann et al. in preparation) and the lack of neutral gas in low-luminosity satellite galaxies of the Milky Way 
(Grcevich et al., 2008).   The HVCs themselves might be the neutral
cores of larger, pressure-supported clouds embedded within this hot gas halo
 (MB04; Collins et al. 2005; Thom et al. 2006; Peek et al. 2007).
 
 Quasar absorption line studies suggest that normal Milky-Way-type galaxies at intermediate redshift are surrounded by extended, $\sim$ 100 kpc, haloes 
of cool gas clouds ($T \simeq  10^4 - 10^5$ K) with high covering factors ($f \sim 0.6 - 0.8$). Less massive 
galaxies tend to have smaller, less pronounced gaseous haloes (Steidel, 1995;  Chen et al., 2001;  Tumlinson \& Fang, 2005; Kacprzak et al., 2008;  Chen \& Tinker, 2008) and the probability that cool halo gas is present may correlate with 
the color of the galaxy (Barton \& Cooke  in preparation).
Interestingly, however, X-ray observations place tight constraints on the nature of 
hot gas around nearby disc galaxies.  Specifically, it must be of a
relatively low density in order to evade X-ray emission bounds (recent limits include
$S_x < 10^{-14}$ erg cm$^{-2}$ s$^{-1}$ arcmin$^{-2}$, Pedersen et al. 2006;
$L_X < 3.8 \times 10^{41}$ erg s$^{-1}$, Benson et al. 2000; $L_X < 3
\times 10^{39}$ erg s$^{-1}$, Li et al. 2007).  These results, together with the fairly high
covering factors in cool clouds implied by absorption line studies
may suggest a picture where
normal galaxies are surrounded by extended, low-density hot ($\sim
10^6$ K) haloes that are filled with fragmented, pressure supported
cool ($\sim 10^4$ K) clouds (MB04; Mo \& Miralda-Escude 1996).

Independently, models aimed at explaining the optical properties of
galaxies have relied increasingly on the idea that extended,
quasi-stable hot gas haloes develop around massive galaxies (Birnboim \& Dekel 2003; Kere{\v s}
et al. 2005; Bower et al. 2006, Croton et al. 2006; Dekel \& Birnboim 2006).  
It is suggested that these hot haloes may be quite susceptible to feedback mechanisms,
which could stabilise the systems to cooling and help explain the
observed bimodality in galaxy properties (Dekel \& Birnboim
2006). It is possible that in massive galactic haloes (a few times $10^{12}$
M$_\odot$) gravitational quenching by clumpy (fragmentary) accretion could provide
the source of energy (Dekel \& Birnboim 2008).

\begin{figure}
\includegraphics[scale=0.4]{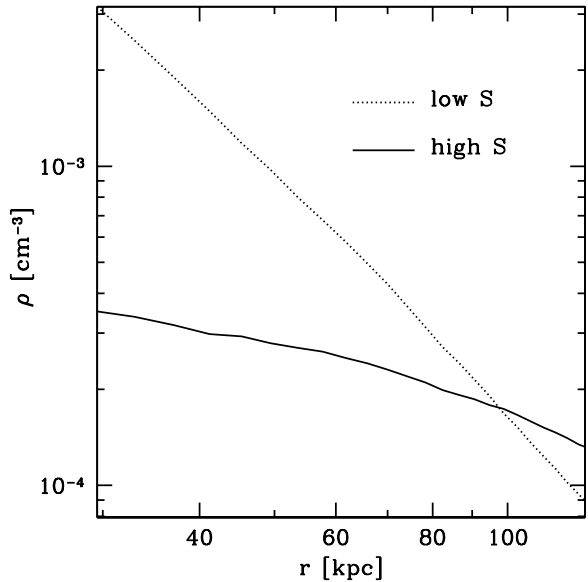}
\caption{Initial gas density profiles for our low and high entropy models.  Note that both models have the same
total gas mass within their virial radii.  The profiles shown have been evolved with an adiabatic equation of state in order
to achieve full relaxation.\label{IC}}
\end{figure}

Observational probes of the gaseous haloes of galaxies provide a
potential means of testing these ideas.  Entropy injected from
feedback mechanisms will alter the density distribution of halo gas
and affect associated cooling rates (and thus X-ray emission) and the
distribution of cooling clouds fragmenting within the hot
haloes. Similarly, early feedback or {\em pre-heating} before the halo
collapses can affect halo gas profiles in a related manner, with
positive consequences for galaxy properties at $z=0$ (Mo \& Mao 2002;
Oh \& Benson 2003; Lu \& Mo 2007).  This type of feedback has the potential
for solving many of the major problems in galaxy formation.
Not only will it help with overcooling, but it may also result in larger
disc galaxies, relieving the so-called angular-momentum problem 
(Navarro \& Steinmetz 2000, Maller \& Dekel 2002).  Moreover,
the X-ray emission in pre-heated haloes is expected to stay
within observational bounds (Mo \& Mao 2002).  The degree to which this type
of pre-heating may affect the fraction of material accreted directly in the form 
cold flows that do not shock-heat (Birnboim \& Dekel 2003; Kere{\v s}
et al. 2005) has yet to be investigated.

\begin{table*}
\begin{center}
  \caption{Simulated galaxies. The fiducial models are shown in the upper
    part, the runs investigating resolution and temperature dependence
    are shown in the lower parts.  \label{table-ic}}
\begin{tabular}{||l|ccccc|lll||}
\hline  \hline 
\, \, \, \, \, (1) & (2) & (3) & (4) & (5) & (6) & (7)\\
\, \, \, Name   & \# gas particles &  $T_{F}$  & Softening Length &
Gas Particle & Dark Matter &  $E_{tot}(t=0)$  \\
\, &       \,   &  [$10^4$ K]  & [kpc] & Mass [$M_{\odot}$] & Particle
Mass [$M_{\odot}$] & [erg]\\
\hline 
low entropy             &$5\times 10^5$     &       3.0   &  0.514 &
$2.8 \times 10^5$ &$2.5 \times 10^6$  & $-3.09\times 10^{59}$ \\
high entropy           &$5\times10^5$     &         3.0   &  0.514 &
$3.6 \times 10^5$ & $2.5 \times 10^6$ & $-2.91\times 10^{59}$ \\
\hline 
low entropy low-resolution       &$1\times 10^5$     &      3.0  & 0.514 &  $1.4\times 10^6$ & $1.3\times 10^7$\\
high entropy  low-resolution             &$1\times10^5$     &         3.0  & 0.514 &  $1.8\times 10^6$ & $1.3\times 10^7$\\
high entropy  high-resolution             &$2\times 10^6$     &        3.0  & 0.257 & $9.0\times 10^4$ & $6.3\times 10^5$\\
\hline
high entropy  LT             &$5\times10^5$     &        1.5  & 0.514 & $3.6 \times 10^5$ & $2.5 \times 10^6$\\
\hline 
\end{tabular}
\end{center}
{\small
(3) $T_F$ is the imposed temperature floor, LT in the name indicates a
low temperature floor. \\
(7) The total energy content of the halo after the evolution with the
adiabatic EOS. }
\end{table*}

In this paper we begin to address some of these questions by simulating the 
formation of a Milky Way sized spiral galaxies starting with two different
initial conditions.  We focus specifically on the cooling of
hot halo baryons -- a process that is expected to be important in fueling galaxy
assembly at $z < 1$ in dark matter haloes of mass $\sim 10^{12} M_\odot$ 
(after the epoch of cold-mode accretion has ended for these objects, Kere{\v s} et al. 2009; Brooks et al. 2009). 
Note that $10^{12} \Msun$ dark matter haloes grow rapidly at 
high redshift, but then grow slowly after $z\sim 1$ (Wechsler et al. 2002; Zhao et al. 2008)
and usually do not experience major ($>1/3$) mergers since that time 
(Maller et at. 2006, Stewart et al. 2008).  Therefore while our simulations do not model the early growth of the galaxy well, they likely provide a reasonable
estimate of late-time cooling flow behaviour.

The first case we consider has a hot gas profile that traces the
density profile of the dark matter.  This, our {\em low-entropy} halo
case, has the lowest central entropy in hot gas that could
realistically be expected.  Our second case explores cooling from a
{\em high-entropy} hot halo, which has a significant core in its
density profile at small radius.  This hot halo has a central entropy
value of $S \sim 30$ keV cm$^2$, which is well within range expected
(and seen) in cosmological simulations that include substantial
non-gravitational pre-heating sources (e.g. Dav{\'e} et al. 2008 find
central entropy values of $S \sim 100$ keV cm$^2$ in small group haloes
of mass $\sim 10^{12.7} M_{\odot}$ as a result of feedback from
outflows).  Another possibility is that dynamical processes could be
responsible for creating hot gas halo profiles of this kind (Kere{\v
  s} et al. 2009; Dekel \& Birnboim 2008; Conroy \& Ostriker 2008;
Khochfar \& Ostriker 2008).  We evolve these two haloes in isolation
without any (additional) form of feedback for 10 Gyr to explore three
questions. One, can large, low-density, hot haloes of the kind
suggested by observations exist in a quasi-stable state for
cosmological time-scales?  Two, can changes to circum-galactic hot gas
at high redshifts persist until today? And three, do different initial
conditions and cooling histories create features that are observable
in galaxies and their gaseous halos today?

 In Section 2, we
present our initial conditions and the numerical techniques. In
Section 3, we discuss the resulting gas halo properties after 10 Gyr of cooling
and discuss these results in the context of gas halo observations.  
The time evolution of the hot haloes as they cool and form central galactic discs are studied in section 4.
Section 5 presents a discussion of numerical convergence.
We conclude and summarise in Section 6.

\section{Smoothed particle hydrodynamic simulations}

We simulate isolated systems with virial mass similar to the Milky
Way.  Haloes are initialised as spherical equilibrium profiles using
the methods outlined in \cite{Stelios}. The virial mass of the model
is $M_{200}= 10^{12}M_{\odot}$ and the dark haloes have NFW density
profiles (Navarro, Frenk \& White 1996) characterised by a halo
concentration $c=8$, where the concentration is defined as
$c=\frac{R_{200}}{R_{s}},$ where $R_{s}$ is the halo scale radius and 
$R_{200} = {206}$ kpc  is the virial radius of the halo (radius corresponding to a density of 200 times the critical density).
Table 1 lists the specific parameters used in each simulation and
provides a reference name for each run.

\subsection{Initial conditions} 
We initialise a fraction of the total halo mass, $f_{\rm b}=0.1$, as a
hot baryonic component with either the same radial distribution of 
density as the
dark matter (the {\em low-entropy} model) or with a flat entropy profile 
that results in a a shallower density profile (the {\em high-entropy} model), 
as shown in Figure \ref{IC}.  Note, that all the models start with the same amount of
baryons within the virial radius of the halo. We then impose a temperature
profile such that the gas is initially in hydrostatic equilibrium for an adiabatic 
equation of state (EOS) where gas cooling is turned off. The high entropy model 
has a central entropy parameter, $S_0 =T_0/n_0^{2/3} \simeq 30$ keV cm$^2$ 
of the type suggested in scenarios with substantial 
pre-heating (e.g. Mo \& Mao 2002).  The amount of entropy in the low entropy model is likely  lower than values arising in cosmological simulations, therefore providing a lower limit in terms of entropy content. However, the assumption that the hot gas follows the radial distribution of the dark matter is widespread in the literature (e.g. van den Bosch 2001).   For all of our 
fiducial models we choose $\lambda_g = 0.03$ for our gas spin parameter, 
defined in analogy with the halo spin as $\lambda_g \equiv j_{g} |E|^{1/2} G^{-1}
M_{\rm 200}^{-3/2}$.  Here, $j_{g}$ is the average specific angular
momentum of the gas, $E$ and $M_{\rm 200}$ are the total energy and mass
of the {\em halo}.

The specific angular momentum distribution of the gas is assumed to
scale linearly with the cylindrical distance from the angular-momentum
axis of the halo, $j \propto r ^{1.0}$.  This choice is consistent
with values found for dark matter haloes within cosmological N-body
simulations (Bullock et al.  2001).  For simplicity, we initialise the
dark matter particles with no net angular momentum.  A detailed
description of our initialisation method is presented in Kaufmann et
al. (2007).

\begin{figure*}
\centering
\includegraphics[scale=0.49]{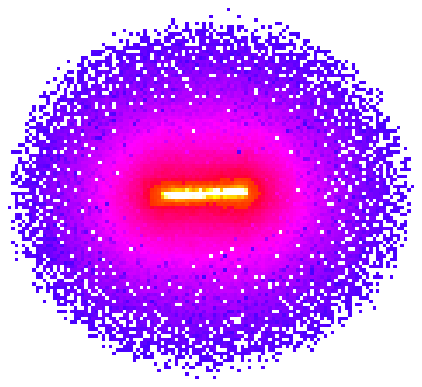}
\includegraphics[scale=0.49]{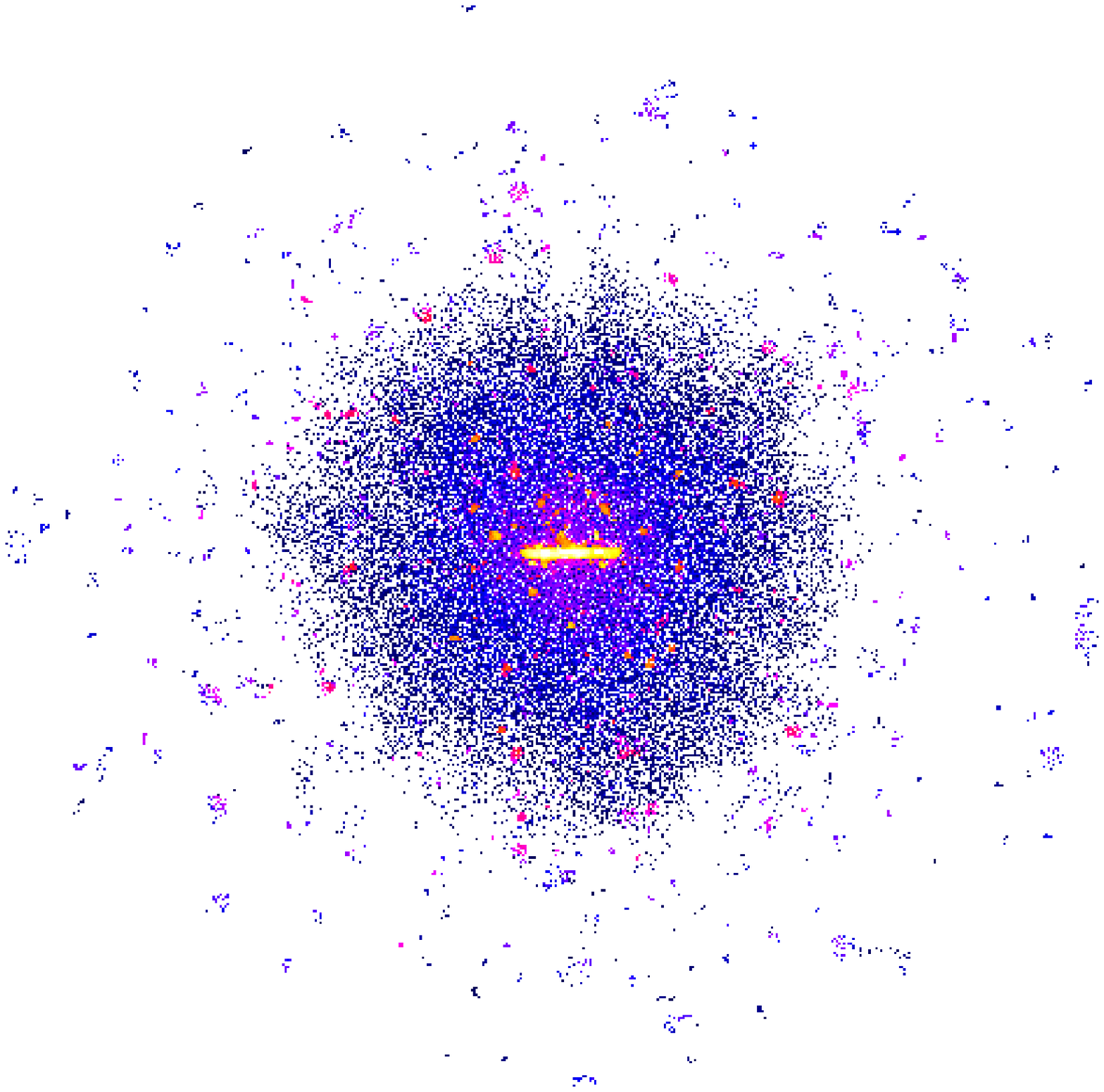}
\caption{Total 3d gas density after 10 Gyr of cooling for the low entropy case (left side) and
the high entropy case (right side). In both cases the box size is  $412$  kpc.  The
distribution of halo gas in the two cases is strikingly different.  In the low entropy case
the gas is highly concentrated within 50 kpc of the disc.  For the high entropy run, the gas is 
spread out much more uniformly at lower density, and dense clumps are clearly seen out 
to radii of $200$ kpc.}
   \label{pic_comp} 
\end{figure*}

Table 1 summarises the models we explore, but our fiducial runs 
use  $N = 5 \times 10^5$ gas and dark matter
particles, and the gravitational softening length is set
to be $0.514$ kpc.  These choices correspond to cases where numerical
losses of angular momentum become small (Kaufmann et al.  2007).
In Section 5, we perform a number of resolution and convergence tests
using lower and higher resolution runs.  The 
low resolution models use $N=10^5$ gas and dark matter particles.
The high-resolution run uses $N = 2 \times 10^6$ 
particles for each of the species and a softening of $0.257$ kpc, scaled as
suggested in Zemp et al. (2008).  The gas and dark matter particle masses 
for the three different resolutions are given in Table \ref{table-ic}.
Finally, we explore two different temperature floors $T_F = 1.5$ and $3 \times 10^{4}$ K.
These values alleviate gravitational instabilities in the disc and
to crudely mimic the effects of 
missing heating sources such as those from an ultraviolet
background (see e.g. Barnes 2002).  They also allow us to explore the properties of 
fragmentary clouds (e.g., their sizes and neutral fractions)
as they are allowed to cool to two different temperatures.
 
Before exploring the effect of cooling in the simulated haloes, we first
allowed them to evolve for $0.5$ Gyr with cooling turned off, in order to allow
the system to fully relax.   Figure \ref{IC} shows the density
profiles after this initial relaxation phase, before cooling is turned on.

\begin{figure*} 
\includegraphics[scale=0.775]{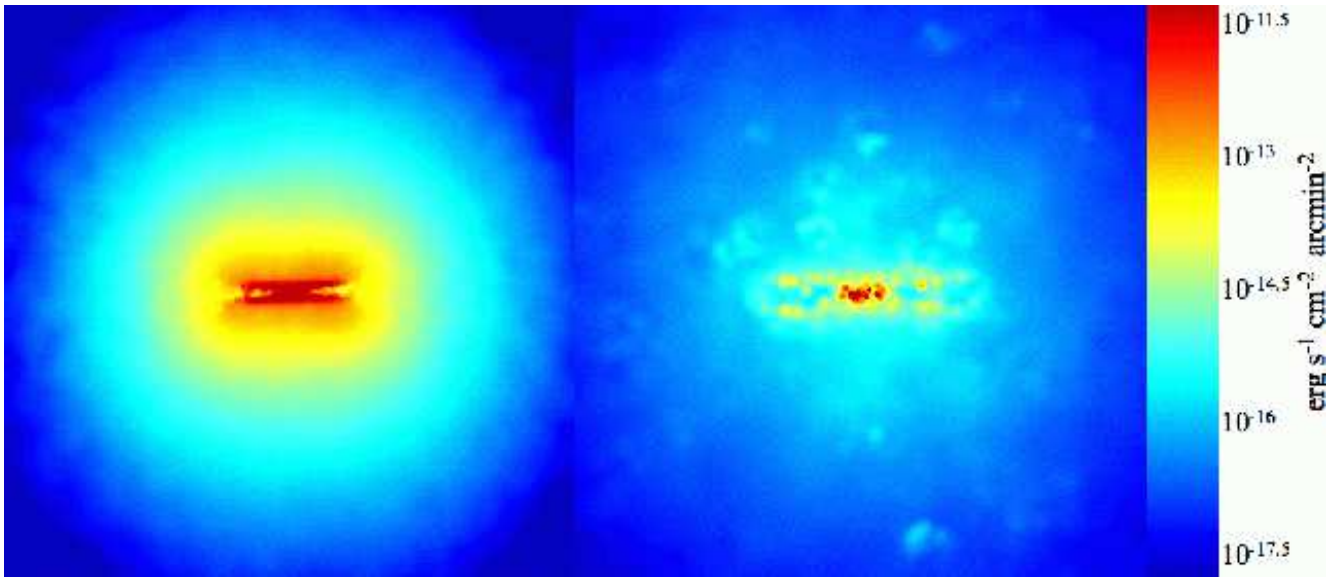}
\centering
\caption{X-ray surface brightness maps of the haloes after 10 Gyr of cooling.  The {\em low-entropy} model is shown on the left
and the {\em high-entropy} model is shown on the right.  Boxes are 100 kpc on a side. \label{mapDHX}    }
\end{figure*}

\subsection{Hydrodynamics and star formation}

We use the parallel TreeSPH (smoothed particle hydrodynamics) code
\textsc{Gasoline} (Wadsley et al. 2004), which is an extension of the
pure N-Body gravity code \textsc{Pkdgrav} developed by Stadel
(2001). It includes artificial viscosity using the shear reduced
version (Balsara 1995) of the standard Monaghan (1992) implementation.
\textsc{Gasoline} uses a spline kernel with compact support for the
softening of the gravitational and SPH quantities. The energy equation
is solved using the asymmetric formulation, which is shown to yield
very similar results compared to the entropy conserving formulation
but conserves energy better (Wadsley et al.  2004).  The code includes
radiative cooling for a primordial mixture of helium and (atomic)
hydrogen.  Because of the lack of molecular cooling and metals, the
efficiency of our cooling functions drops rapidly below $10^4$ K.  The
lack of molecular cooling is unimportant in our investigation because
we enforce temperature floors $T_{\rm F} \ge 1.5 \times 10^4$ K.

 The adopted star formation recipe is similar to that described in
 Katz (1992); stars spawn from cold, Jeans unstable gas particles in
 regions of converging flows. The mass of gas particles decreases
 gradually as they spawn more star particles. Once a gas particle is
 eligible for spawning stars, it does so based on a probability
 distribution function with a star formation rate parameter that
 can be tuned to match the Kennicutt (1998) Schmidt Law.  The mass of the 
 gas particles decreases gradually as they spawn more star particles.
 After its mass has decreased below $10\%$ of its initial value the
 gas particle is removed and its mass is re-allocated among the
 neighboring gas particles.  Up to six star particles are then created
 for each gas particle in the disc. For the fiducial and high
 resolution simulations we allow only one star to spawn per
 gas particle, thus alleviating the computational load. The subsequent
 formation of stars has no (energetic) effect to the surrounding gas,
 i.e., there is no feedback associated with star formation.

\subsection{Time evolution and cloud formation}

As discussed below, 
at some point after cooling is turned on in each halo (the time depends on the
initial condition),  a disc 
forms in the halo centre. In the low-entropy case, the disc is built from gas that cools rapidly near the halo centre.
In the high-entropy case, the disc is built primarily from cool $T \sim T_F$ clouds
that condensate within the extended hot halo and fall in to the disc region on $\sim 2$ Gyr time-scales.

The cool clouds that form in the high-entropy halo arise from a
physically expected phenomenon, the thermal instability (most likely as described in Field 1965; see also MB04 and Kaufmann et al. 2006).  The thermal
instability has less time to develop in the low-entropy halo because
the cooling flow is so rapid towards the central region in that
case\footnote{One of the conditions for cloud formation is that
    the sound-crossing time, $\tau_{\lambda}\simeq\lambda_{i}/v_{s},$
    across a perturbation of wavelength $\lambda_{i},$ should be less
    than the characteristic cooling time (MB04). If this condition is
    not satisfied the perturbation is erased because the local cooling
    time is too close to the mean cooling time and the whole region
    becomes isothermal. While roughly the same level of perturbations
    are resolved in the low and high entropy simulations and the sound
    speed ($v_s\propto \sqrt{T}$) remains comparable, the cooling time in
    the centre of the low entropy halo is much shorter due to the
    higher gas density as well as a lower initial temperature and its
    perturbations get erased. The cooling time in the high entropy
    case is also longer because the cooling rates for primordial gas
    drop from $T\sim 4\times 10^6$ K to $\sim 10^6$ K (Katz et al. 1996),
    whereas the low entropy halo starts in the rising part of the cooling curve, see Figure
    \ref{dens_comp}.}.

However, the fluctuations that seed the thermal instability in our simulations are unphysical.  
Specifically, the fluctuation field is numerical Poisson noise, and
tracks the initial density fluctuations in the SPH particle 
distribution.   Though we  expect that real galaxy haloes will have fluctuations in the hot gas distribution,
it remains difficult to predict the expected fluctuation field self-consistently from first principles (see MB04).
In this sense, our initial (Poisson) fluctuation spectrum may be regarded as an exploratory step towards 
understanding this complicated problem.

Encouragingly, as we discuss in Section 5, there are many global properties of the simulated gaseous haloes themselves that are
 robust to the input noise field.  While the mass spectrum of fragmentary clouds is sensitive to the initial perturbation spectrum,
 the integrated mass in cool halo clouds is almost invariant to it.
Moreover, the hot gas mass and density profile; the total resultant disc mass; and  the time evolution of the system is
convergent.  

Most importantly, the gross difference between the high-entropy halo (which produces clouds) and the low-entropy halo (which does not)
 is also a physical difference -- the initial fluctuation amplitudes 
 in the high-entropy and low-entropy haloes are
 very similar (Section \ref{reso}).    We defer the somewhat lengthy discussion of
the cloud property dependence on resolution and on other physical processes in
Section \ref{reso}.

\begin{figure*} 
\includegraphics[scale=0.775]{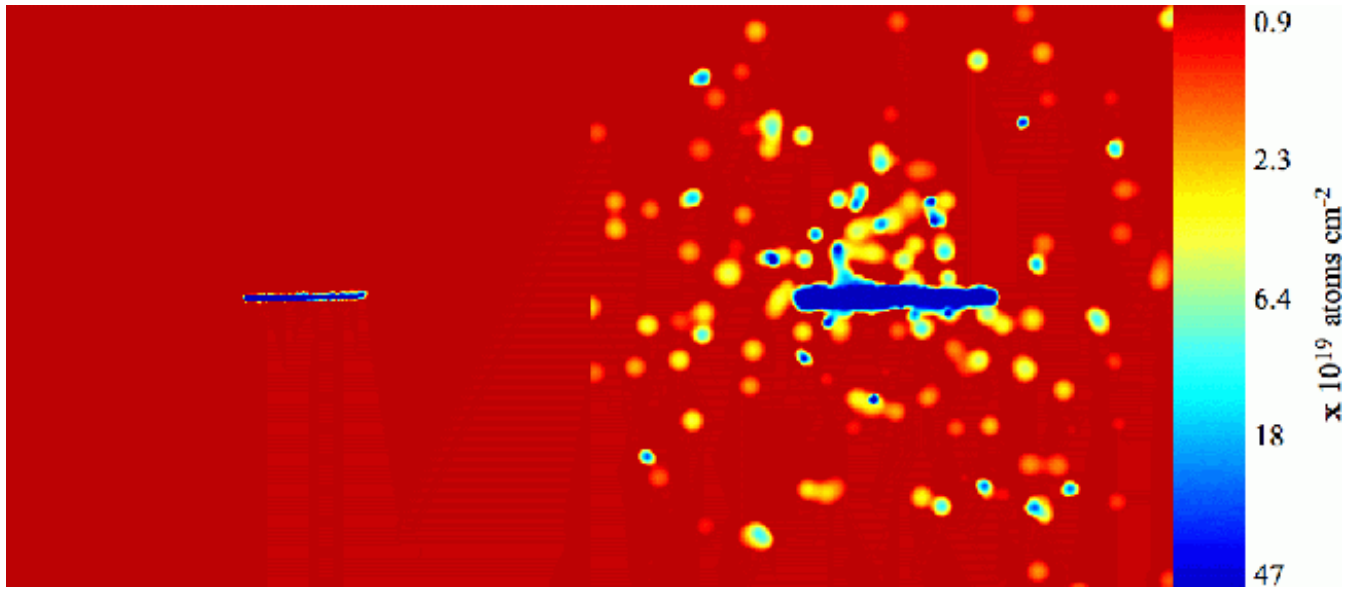}
\caption{Maps of the projected HI density  for the low-entropy case (left) and high entropy case (right).
Shown is the case for $T_F = 15\,000$ K, see text for details. Boxes are 100 kpc on a side. \label{mapH}} 
\end{figure*}

\begin{figure*} 
\includegraphics[scale=0.775]{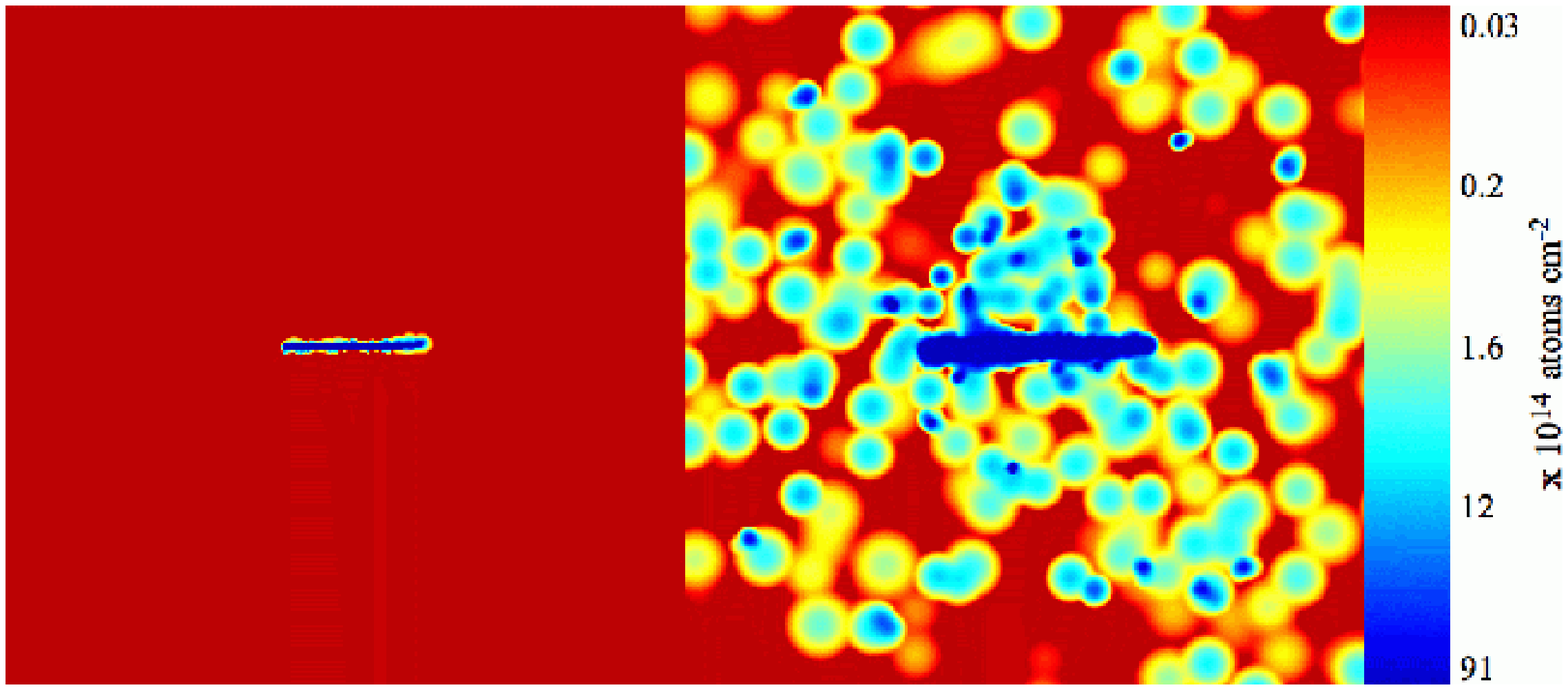}
\caption{Maps of the projected \mg density  for the low-entropy case (left) and high entropy case (right).
Shown is the case for $T_F = 15\,000$ K and 0.3 metallicity gas, see text for details. Boxes are 100 kpc on a side. \label{mapmg}}
\end{figure*}

\begin{figure} 
\includegraphics[scale=0.6]{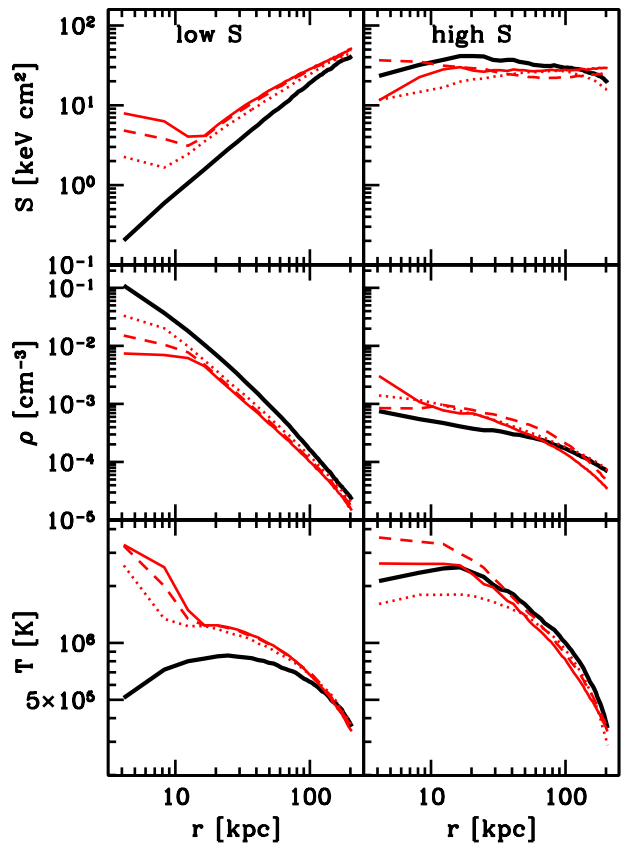}
\caption{\label{dens_comp}Evolution of entropy (top), density (middle)
  and temperature (bottom) of the hot gas for the low entropy (left)
  and high entropy (right) models.  The lines represent 0 (thick
  solid), 3 (dotted), 7 (dashed) and 10 (solid) Gyr of cooling. In the
  low entropy case the entropy increases in the centre of the halo as
  the lowest entropy material cools out and the average temperature of
  the hot gas increases. The density profile monotonically decreases
  and a small core in the hot gas starts to form.  In the high entropy
  case the entropy, the density and the temperature oscillate around
  the original values with little evolution. \emph{A hot gas halo that
    has this profile can remain in a steady state for 10 Gyr.}}
\end{figure}

\begin{figure*} 
\includegraphics[scale=0.8]{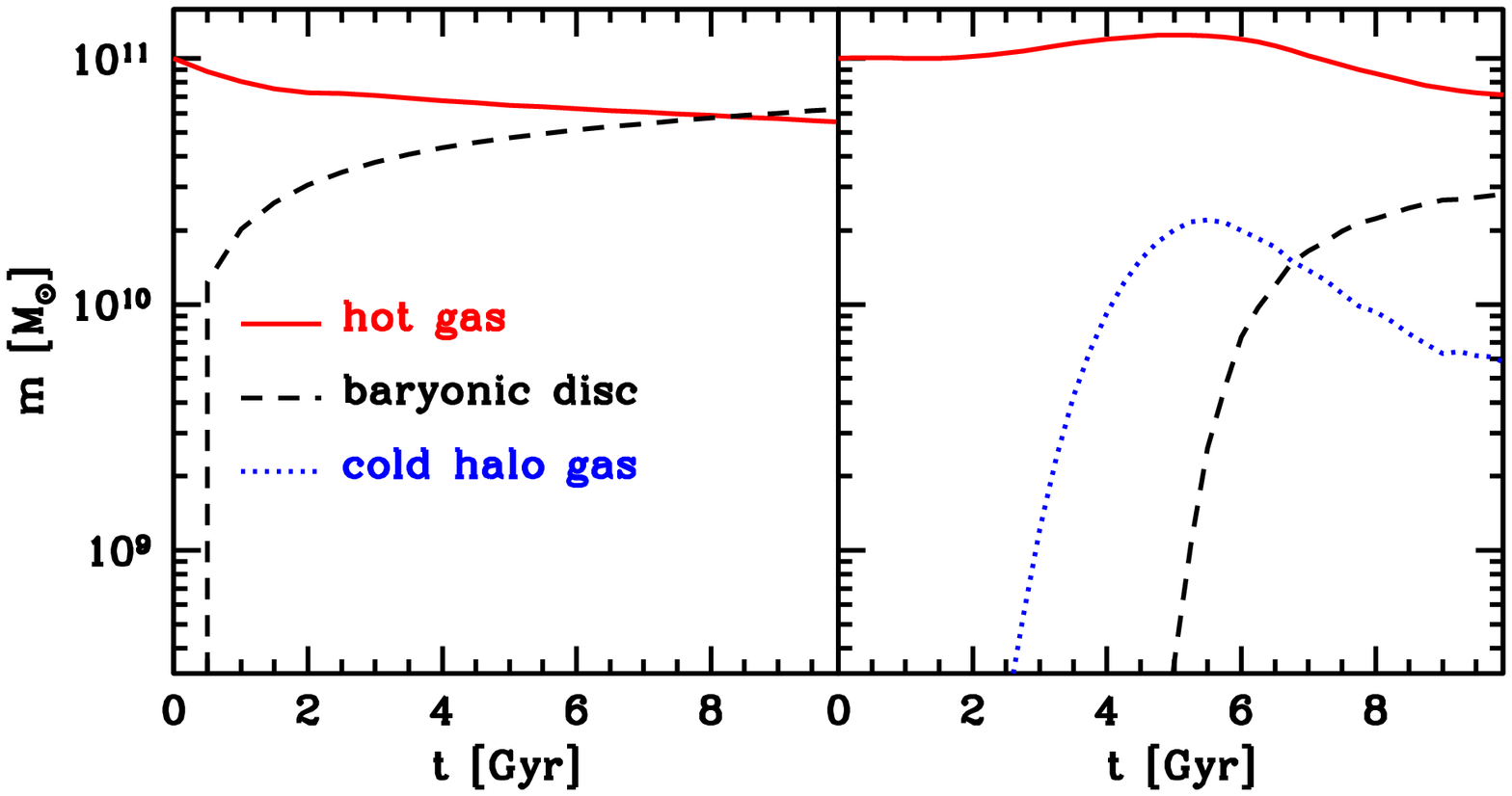}
\caption{Mass evolution of the different gas phases, {\bf left} low entropy and {\bf right} high entropy model. Note, that the low entropy model does not show a significant amount of cold halo gas.}
  \label{massevo_comp}
\end{figure*}

\section{The resulting halo gas}
\label{reshalo}
First we focus on the simulation results after 10 Gyr of cooling.
 Figure \ref{pic_comp} shows images of the 3d gas density for each case,
 and illustrates that the two scenarios create strikingly different gas halo distributions.
In the left panel, we see that the low-entropy halo has cooled to form
a central disc and that the gas density is quite high only in the central disc region.
  The high entropy halo (right) has produced a disc that is less massive
  (see below) along with an extended 
$\sim 200$ kpc distribution of fragmented, cool $T = T_F \sim 10^4$ K 
clouds.   The total cool mass that resides
in the final high-entropy halo is 
 $\sim 5 \times 10^{9} \, \Msun$ after $10$ Gyr of evolution (see Table 2). In contrast,
the 
low entropy case (left) yields virtually no extended distribution of cool clouds 
in the halo.  
It is clear then that the initialised differences in the gas profiles persist and result in
striking differences after $10$ Gyr of cooling.
 This suggests that observations of 
the gaseous haloes of galaxies can be used to understand their cooling 
history. Below we discuss these results in the context of observational signatures.

\subsection{X-ray observations}

Figure \ref{mapDHX} shows X-ray surface brightness maps for each halo
after 10 Gyr of cooling.  The X-ray emissivities were calculated in
the 0.1 to 10 keV band using the MEKAL software package\footnote{see
  http://heasarc.gsfc.nasa.gov/docs/xanadu/xspec/} assuming a hot gas
metallicity of $0.3$ solar.  As mentioned in the section \ref{sec:intro}, some observations suggest an upper limit for X-ray emission from disc galaxies of
$S_x < 10^{-14}$ erg cm$^{-2}$ s$^{-1}$ arcmin$^{-2}$  (Pedersen et
al. 2006).  The emission of the hot halo in the low entropy run is up to 
$\sim 100$ times too bright in the X-ray, while the high entropy case
is within these observational bounds. 

Benson et al. (2000) reported limits on the X-ray luminosities of three galaxy halos of $L_X <$ 0.4, 1.2 and 3.8 $\times 10^{41}$ erg s$^{-1}$.
In this investigation, the authors did not include luminosity from the central disk region in order to concentrate on halo gas explicitly.
In Table 2 we list the $L_X$ values for each of our halos, excluding the central 10 kpc region by analogy with Benson et al. (2000).
The luminosity from the the high entropy model ($L_X = 2.2 \times 10^{39}$ erg s$^{-1}$) is well below the observed limits, while the value for the low entropy model  ($L_X = 1.3 \times 10^{41}$erg s$^{-1}$) is inconsistent with two of the halo $L_X$ limits from Benson et al. (2000).

Finally we note that even our high-entropy halo is too bright in the X-ray in the very central disk region.  Unfortunately, the predicted 
luminosity in this region suffers
from both numerical limitations and from the lack of important feedback physics, which almost certainly is important on this scale. 
 First, at the disc-halo
interface, the densities of the hot gas particles and therefore their X-ray
emission are over-estimated due to the SPH smoothing
procedure\footnote{However, one can show that if the cold gas
  particles are cut away and the densities of the remaining gas
  particles are then recalculated, the X-ray luminosities will be
  underestimated (Toft et al. 2002)}. Furthermore, feedback from
star-formation is expected to modify the thermal structure of the hot
halo at the disc-halo interface and this could act to lower the cooling rate there.  
If we blindly calculate the X-ray luminosity that comes from the central $10$ kpc region we find $L_X (<10 {\rm kpc}) = 2.9 \times 10^{41}$ erg s$^{-1}$ for the high entropy model and  $2.0 \times 10^{43}$ erg s$^{-1}$ for the low entropy model.  The latter is almost four orders of magnitude higher than the measurement of Li et al. (2008) who find a total X-ray luminosity from the central part NGC 5775 to be $\sim 3.5 \times 10^{39}$ erg s$^{-1}$. 
Realistic comparisons to these observations must be deferred until more accurate simulations are run, but it would seem difficult to
reconcile a factor of ten thousand disagreement, as is seen with the low-entropy model.

\subsection{Cool-gas observations}

As discussed in Section 5, the
cloud mass spectrum is sensitive to the initial fluctuation field, therefore our
results in this section should be regarded as an exploration of the expected trends rather
than detailed first-principle predictions.  Nonetheless, we expect the differences
between the high-entropy and low-entropy cases to be robust.

The \hi\,HVCs in the Galactic halo are potentially building up the
baryonic mass of the Galaxy (Putman 2006). To compare with
observations of the local HVC population, we derived the neutral \hi
\,fraction of the cold material using Cloudy (version C06.02), last
described by Ferland et al. (1998). The chosen radiation field will influence
the \hi \, fraction heavily but several observations suggest that
extragalatic ionizing photons are likely the source of the ionisation
(see e.g. Maloney 1993), so we adopt the metagalactic background
spectrum of Sternberg et al. (2002), which covers wavelength from
infrared to X-rays. The ionizing photons are mainly produced by
quasars and star-forming galaxies. The influence of radiation from the
Galaxy is smaller: Assuming a small but very uncertain escape fraction
of the Lyman continuum flux (contributed mainly by the OB
associations), {$f_{esc}\sim 5\%$ } (e.g., Dove, Shull, \& Ferrara
2000), the metagalactic contribution dominates over the Galactic one
at a distance of $\sim 180$ kpc. For soft X-ray photons, the two
contributions equal at $\sim 16$ kpc (Slavin, McKee, \& Hollenbach
2000). A variation of the ionizing flux by a factor of 2 may affect
the estimation of the gas mass by a factor of 1.5 (Maloney \& Putman
2003).

The amount of neutral \hi \, in the simulations depends also
critically on the temperature of the cold clouds (which have cooled
down to the temperature floor set in the simulation).  The $30\,000$ K
temperature floor prevents the formation of a large amount of neutral
\hi, the large core model predicts $\sim 1.5 \times 10^{6} \, \Msun$
of neutral \hi \,distributed in the halo (out to $200$ kpc).  But
(background) radiation from stars and cosmological sources might not
be able to keep the gas as hot. We checked that dependence by evolving
the two fiducial models further for $40$ Myr with a temperature floor
of $15\,000$ K; the neutral \hi \,mass in the halo is then $\sim 9
\times 10^{8} \, \Msun$ for the high entropy model: this is an increase of more
than a factor $100,$ which demonstrates how sensitive the \hi\,fraction is to the
temperature of the cool clouds.

The high entropy case with a temperature floor of $15\,000$ K is in
fairly good agreement with the observations of the Milky Way. Putman
(2006) estimated the total gas mass in the HVCs of the Milky Way to be
$1.1 - 1.4\times10^9\ M_{\odot}$, if assuming the clouds are
distributed within 150 kpc. Our prediction is somewhat higher
($5.5\times10^9\ M_{\odot}$, Table \ref{table-disks10}), but the
estimated total gas mass in the observed HVCs would increase if their
average distance is somewhat higher and/or their \hi \, fraction is
lower than assumed (Putman 2006). The velocity field of the simulated clouds is quite similar to the one of observed HVC's, but since it is dependent on the initial spin of the hot gas and initial noise field, we defer a detailed comparison.

The choice of $15\,000$ K for the high entropy model also agrees quite
well with observations of the \hi \,mass in the nearby massive spiral
NGC 891, for which Sancisi et al. (2008) quote an \hi \, mass in the
halo of $\sim 1.2 \times 10^{9} \, \Msun$.  Those authors also infer a
mean ``visible'' accretion rate of cold gas in galaxies to at least
$0.2\, \Msun \, yr^{-1}$; today's rate for the high entropy model can
be estimated by multiplying the total mass accretion rate to the disc
at the final time step ($2.05\, \Msun yr^{-1}$) with the fraction of
cold accretion ($0.75$) and the \hi \,fraction ($0.18$). The visible
accretion rate in the large core model is therefore $\sim 0.28 \,\Msun
\, yr^{-1}$.  Rand \& Benjamin (2008) found less \hi \,($\sim 10^{8} \,
\Msun$) around NGC 5746, a edge-on galaxy known to have a large
corona. The radiation field around that galaxy might be somewhat
higher, preventing some \hi \,to form, and the cool clouds stay more
ionised as expected in the case of a temperature floor higher than
$15\,000$ K.  The covering fraction for \hi \,with number densities
$N(\hi)> 2 \times 10^{18} \, {\rm cm}^{-2}$ is approximately $35\%$ in
the high entropy simulation similar as that found around the Galaxy
(Murphy, Lockman \& Savage 1995). 

The low entropy model does not show any neutral \hi \,outside of the
disc for either temperature floor.  While Figure \ref{mapH} illustrates the
situation with the lower temperature floor, the derivation
of self-consistent \hi \, masses would require a simulation, which
follows the feedback from the radiation field from stars and
background sources self-consistently. The values given above are
likely bracketing the expected \hi \,masses but they also show, that
the \hi \,clouds are always embedded in substantial clouds of ionised
hydrogen. Furthermore, a galactic fountain - not yet included in the
simulations - might produce more extra-planar gas (Fraternali \&
Binney 2008).

We also calculate the \mg \,ionisation fraction using CLOUDY, and
estimate the covering fraction of the \mg\,gas clouds (see the maps in Figure 5 and summary
in Table \ref{table-disks10}). Again, the covering fraction given by the high
entropy case with temperature floor of $15\,000$ K, $\sim 0.6$, is
consistent with observations. For $\sim 20$ galaxies, Tripp \&
Bowen (2005) found a \mg\,covering fraction of $\sim 0.5$, whereas Chen \& Tinker (2008) report a \mg\,covering fraction of $\sim 0.8$. Bechtold \&
Ellingson (1992) reported a smaller cover fraction, $\sim 0.25$, but
they observed within a bigger impact distance, $D \leq 85$ kpc.

\bigskip 

In summary,  the high entropy model can reproduce almost all of the
observed qualitative features of halo gas.  The differences between the high and low
entropy models in their X-ray luminosities are particularly striking and should
be robust to numerical uncertainties.

The high-entropy model is also able to reproduce observed covering
fractions for absorption systems.  However, the predictions
depend sensitively on the temperature floor adopted (which acts as a proxy
for the external ionizing field) and, as mentioned earlier, on the initial (unphysical) noise field
in the simulations.  Nonetheless, for the same initial noise field, the high-entropy model facilitates the formation
of fragmentary clouds while the low-entropy model does not.
Specifically, the low-entropy model has no  cool halo clouds  to be seen as
high velocity clouds or quasar absorption systems, while the high-entropy model allows the clouds
to emerge over time-scales that are longer than the inflow time-scale associated with
cooling from the central region.

\section{The time evolution of the hot gas halo and the galactic disc}

Figure \ref{dens_comp} illustrates the evolution of the hot gas ($T> 50\, 000$ K) density profiles (middle), temperature (lower) and
entropy profiles (upper) for each of our initial haloes (the low and high entropy
models are shown in the left and right panels, respectively).
It is clear that the different initial gas 
distributions lead to a completely different cooling behaviours. The 
low entropy halo cools quickly from the central region.
Over the same period, the hot gas 
density of the high entropy model remains remarkably stable and the
entropy profile remains quasi-stable, but does experience some
oscillations.  We show in Section 5 that these results are stable to
numerical convergence.

The build up of the disc and cold gas populations in both scenarios can be seen in Figure 
\ref{massevo_comp}, which shows the mass evolution in the central disc galaxy (dashed),
and the hot halo component (red, solid) as a function of time.  The hot gas includes
all $T> 50,000$ K gas within the halo virial radius.  
  In the low entropy case, the hot gas quickly cools within 
a $\sim2$ Gyr to form a massive disc ($2\times 10^{10} \Msun$). Then cooling
proceeds at a slower rate with the hot halo being depleted and the disc growing 
larger to $6 \times 10^{10} \Msun$ after 10 Gyr.
 In contrast, in the high entropy case, no cooling occurs for the first 
$\sim 2$ Gyr.  Then the hot gas starts to fragment into cool clouds that are embedded within
the hot halo (blue  dotted line) which eventually fall in to form a central disc\footnote{The time evolution beyond 10 Gyr showed,  that the density profile in the high entropy case remained relatively stable  and the mass growth of the galactic disc slowed down compared to the period from 5 to 10 Gyr. }.  Note that
the halo cloud population peaks with a total mass of $2\times 10^{10} \Msun$
 before a central disc starts to form. The hot gas in this case shows a slight increase in mass at $\sim3$ Gyr as hot gas
enters the halo from beyond the virial radius. A similar upturn is not seen in 
the low entropy case because the gas always cools faster than new gas enters the halo virial radius.

The final disc in the high entropy case is significantly less massive
$\sim 3 \times 10^{10}$ $\Msun$ and also has $\sim 60 \%$ more
specific angular momentum than in the low entropy run (see Table
\ref{table-disks10}). The reason for this is that the disc is build up
not only from gas cooling close to the centre of the halo and getting
incorporated to the disc but also from material which cooled already
far away from the central disc. Roughly $45\%$ of the material that
makes the disc in the high entropy halo is accreted as cool clouds
that formed beyond $30$ kpc from the disc. Note that, while the total
angular momentum of the gas in the two simulations was the same
initially, the discs form out of different portion of the specific
angular momentum distribution.

\begin{figure} 
\includegraphics[scale=0.35]{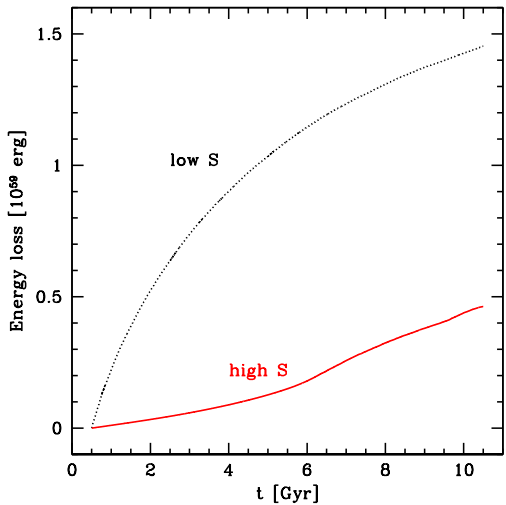}
\caption{The time evolution of the total energy loss of the system is
  shown. Cooling is much more efficient in the low S model compared to
  the high S simulation. \label{energy} }
\end{figure}

These trends can be clearly understood by looking at the energy of the 
hot gas (Figure \ref{energy}). The low entropy run loses most of its
energy early on while the energy losses of the high entropy run are quite
modest over the $10$ Gyr.   Note that, although
the low and high entropy runs are losing energy at a similar rate from 6 Gyr 
on, their X-ray surface brightness emission (Figure 3) is very different due to different spatial distribution
of the cooling --- the low entropy run cools in the centre whereas cooling in 
the high entropy run is diluted in the whole halo.

Interestingly, the radial scale-length and the specific angular momentum of the disc after $10$ Gyr
is larger in the high entropy run than in the low entropy run.  This
shows, that even though the cool clouds might lose angular momentum
due to ram pressure while sinking to the centre, this path of disc
formation actually increases the angular momentum content by more than
$50 \%$ compared to the low entropy disc (Table \ref{table-disks10}), possibly
relieving the ``angular momentum problem'' in disc formation.

Besides producing a gas halo that agrees with observations it appears that 
the high entropy simulation is also able to create a disc galaxy that does 
not suffer from the common problems associated with disc galaxy formation 
in numerical simulations.  The reduction in disc mass and increase in specific angular 
momentum  suggests that 
building discs from cool clouds is a promising path.

\begin{table}
\begin{center}
  \caption{Various results for the fiducial models  after $10$ Gyr  \label{table-disks10}}
\begin{tabular}{||l|cccc|lll||}
\hline  \hline 
\, \, \, model & high S   &  low S   \\

\hline 

(1a)$^*$ \hi \, cloud mass in $\Msun$ at $15000 $K &  $9 \times 10^{8} $      &  $0$ \\
(1b)$^*$ \hi \, cloud mass in $\Msun$ at $30000 $K  &  $1.5 \times 10^{6}$     &  $0$ \\
(2a)$^*$ \hi \,covering fraction at $15000$K & $0.36$         & $0.01$  \\
(2b)$^*$ \hi \,covering fraction at $30000$K  & $0.00$         & $0.00$  \\
(3a)$^*$ \mg covering fraction at  $15000$K & $0.60$         & $0.01$  \\
(3b)$^*$ \mg covering fraction at $30000$K  & $0.02$         & $0.00$  \\
(4) total cold halo gas in $\Msun$ &  $5.5 \times 10^{9} $  &  $0$  \\
(5) total hot halo gas in $\Msun$ & $7.1 \times 10^{10}$ & $5.5 \times 10^{10} $ \\
(6) total disc mass in $\Msun$   &  $2.9 \times 10^{10}$ & $6.3 \times 10^{10}$ \\ 
(7) disc scale length in kpc            &  $3.6$  &  $2.6$ \\
(8) disc j in kpc km s$^{-1}$         &  $740$   & $468$  \\
(9) $L_X$ in erg s$^{-1}$ & $2.2 \times   10^{39}$ & $1.3 \times   10^{41}$\\
\hline 
\end{tabular}
\end{center}
{\small

(1) Total \hi \,mass in the halo, see section 3.2.\\
(2) Covering fraction for $N(\hi)> 2 \times 10^{18} \, cm^{-2}$ within
a box of 100 kpc for,  see section 3.2.\\
(3) Covering fraction for $N(\mg)> 10^{13} \, cm^{-2}$ within
a box of 100 kpc for,  see section 3.2.\\
(4) Cold gas within the virial radius, without the gaseous disc. \\
(5) Hot gas within the virial radius. \\
(6) Total disc mass including gas and stars \\
(7) Radial scale length of the stellar disc measured by
    fitting an exponential to the surface density profile neglecting the
    bulge region.\\
(8) Specific angular momentum of stars plus cold gas disc. \\
(9) Bolometric X-ray luminosity within a 100 kpc box around the disc for a metallicity of 0.3 solar. A 10 kpc box around the centre of the disc has been cut out, see section 3.1.\\
$^*$  Number so marked is dependent on the initial noise field in the simulation and is presented to guide expectations and to
contrast results between the two initial setups. \\
 }
\end{table}

\section{Resolution Tests and Exploration of Cool Cloud Properties}
\label{reso}

We have conducted resolution tests in our high-entropy halo by increasing the number of gas particles  within the virial radius
from $10^5$ to $2 \times 10^6$ in a series of
three simulations (see Table \ref{table-ic}). While   
the cool clouds that form in these simulations arise from a real, physical  process that amplifies initially
small perturbations (the thermal instability)
the initial seed perturbations are not physically motivated.  Specifically, they are set by Poisson sampling noise (Kaufmann et al. 2006).  
Therefore, the noise field that arises from our sparse sampling statistics may be regarded as a toy model
example of the fluctuation field that is expected to exist in real galaxy haloes (see MB04 for a discussion).
Unfortunately,  the true nature of the expected perturbation spectrum will likely remain difficult
to predict from first principles for many years.  This motivates us ask whether there are quantities that remain invariant
to different perturbation distributions.

We would like to determine whether our ignorance associated with the expected perturbation field prohibits us from
predicting any generic features of the cooling haloes.  As we now describe, we find perhaps the most encouraging result
possible: most global properties of the resultant
haloes (including the integrated mass in cool halo material and the radial extent of this cool halo gas) are
 robust to the initial noise field.  The initial fluctuation spectrum seems to affect only 
the resultant cloud mass spectrum.  Of course, the latter fact implies that the cloud mass spectrum and
related quantities like cloud covering factors are subject to very large uncertainties.  There is hope that
the `real' fluctuations can be modeled effectively in very high resolution cosmological simulations, but disentangling these expected
fluctuations from numerical noise will be a difficult task.  Even in cosmological simulations,
 it may be best to concentrate on robust, global quantities, like the integrated mass, at least in the near future.

\begin{figure}
\includegraphics[scale=0.35]{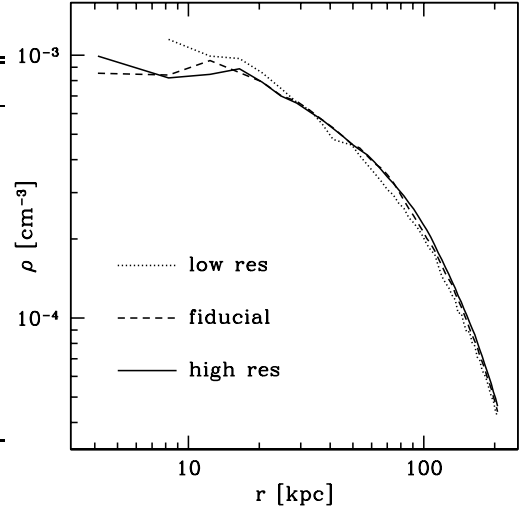}
\caption{The hot gas density of the high entropy model after 7 Gyr of
  evolution at different resolutions: The profile remains basically unchanged for  an increase of a factor $20$ in particle number. \label{profile_resol}}
\end{figure}

Figure \ref{profile_resol} shows that the hot gas density profile of the the high entropy model is convergent
after $7$ Gyr at different resolutions.   Therefore, the fact that  a quasi-stable hot gas halo can exist
over a significant fraction of the Hubble time  holds at all the resolutions tested.
Moreover, the integrated mass and radial extend of cold halo gas (the total mass in halo clouds) is also convergent
over these three cases, as summarised in Table \ref{table-conv}.  We also have run one lower resolution simulation for the
low-entropy case (with $10^5$ gas particles, see Table 1) and find that the global results for this case are almost identical to
the higher resolution run.

\begin{figure} 
\includegraphics[scale=0.35]{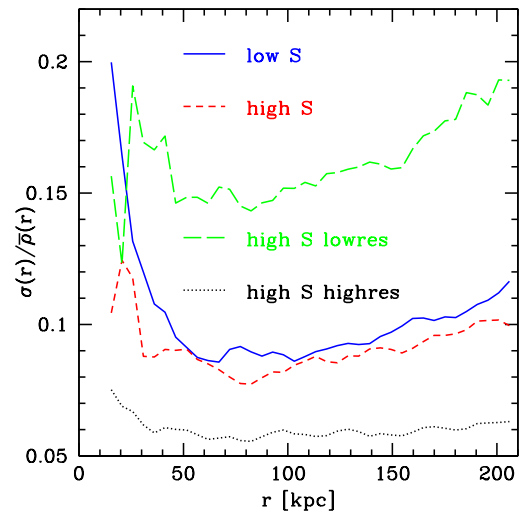}
\caption{Comparison of the perturbations in the gas density after the
  evolution with an adiabatic EOS. The RMS density fluctuation divided by the mean density in that radial bin is plotted
  versus radius  (the local density was smoothed over the same mass for the different resolutions).  \label{noise} }
\end{figure}

\subsection{Cloud Properties: Resolution and Temperature Floor Tests}
  
We identify clouds using a friends of
friends algorithm (FOF) with $32$ particles as the minimum threshold (equal to the number of
particles in the smoothing kernel).   The linking length was chosen to be 0.2 times the mean particle separation, 
and we checked in the standard simulation that the number
of clouds found with the FOF depends only weakly on the exact
choice of the linking length.
We note that we have additional diffuse, cool material in the halo
 that cannot be identified or resolved 
by our 32 particle FOF algorithm.  This additional material is included in our accounting of total `cool halo gas'.

We expect that the resultant 
cloud mass spectrum will depend on the
level of initial density perturbations, and hence on the resolution. Figure \ref{noise} shows
an illustration of the initial RMS density fluctuations as a function of radius
for the standard-resolution high-entropy and low-entropy runs before cooling is turned on
(i.e. after the $0.5$ Gyr of evolution with an adiabatic
EOS).  Specifically, we
measure the RMS density variation around each particle at radius $r$
using the nearest $n=32$ particles (the number used in the smoothing kernel of the simulations) to calculate its local density, compared to the mean density averaged in the respective radial bin.
We see that the noise level is comparable between the two simulations (and even slightly larger in the
low-entropy case).   This demonstrates that it is not a difference in the
initial
perturbations that gives rise to the formation of clouds in the high-entropy run, but rather the 
difference in the initial density profiles. 

\begin{figure} 
\includegraphics[scale=0.35]{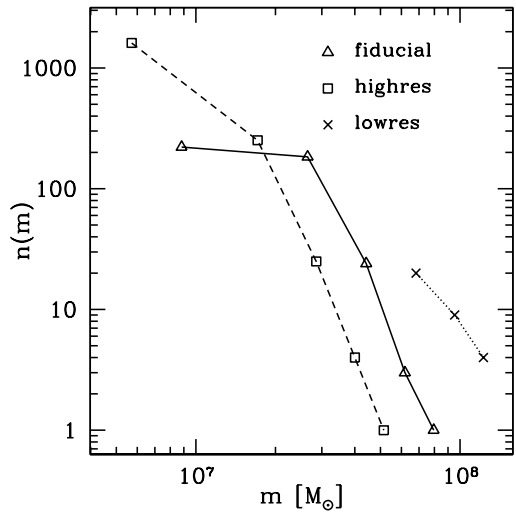}
\caption{Cloud mass spectra for the high entropy run at different
  resolution (fiducial, low-res and high-res) after $6$ Gyr of evolution with cooling. Note, that even the most massive clouds in the low-res run do not contain more than a hundred particles.
   \label{cloudmasses}}
\end{figure}

 Figure \ref{noise} also shows the fluctuation
amplitude for the lower and higher resolution realisations for the high-entropy case (where here
the value of $n$ in the measurement is adjusted to $n=6$ and $n=128$ respectively so that the total mass averaged over remains fixed).
The noise level varies as expected from Poisson statistics.  
The resulting cloud mass functions for the three high-entropy runs are shown in Figure \ref{cloudmasses}.
We find that when the initial 
fluctuation amplitude is smaller (from higher particle number) the spectrum of cloud
masses is correspondingly reduced (as $\sim 1/\sqrt{N}$ in the particle number).
Thus any prediction that depends on the mass spectrum of clouds, including
cloud sizes and covering fractions, will require an understanding of the initial perturbation spectrum.
We note that it has been shown
(Kaufmann et al. 2006, their Figure 12) that if the level of perturbations is kept fixed while
increasing particle resolution, the resultant cloud spectrum is independent of particle number (down to
the resolution limit) therefore the noise spectrum itself is the critical unknown quantity in these simulations.  
Though the mass spectrum of clouds does not converge, we
find that the total cool mass in the halo and its radial extent in the large
core simulation is fairly independent of the numerical resolution,
even for an increase of the number of gas particles by a factor of
$20$, see Table \ref{table-conv}.

We find that global differences due to using a different temperature floor are  small.
Clouds do cool down to the respective temperature floor imposed,
so heating by the clouds motion through the hot background is either
not important or not captured by the code. The spatial and the velocity 
distribution of the clouds after $7$ Gyr in the high entropy LT run with 
$T_F =15\,000$ K are comparable to the fiducial run at $T_F = 30\,000$ K. 
On the other hand, the densities of the
clouds in the LT run are increased due to the smaller internal pressure, as
expected. Nonetheless, the exact value
of the lowest temperature imposed is not of critical importance for the total mass in clouds.
However, it is very important for the \hi \,mass fraction in each cloud (and in the expected covering fraction)
as discussed in Section \ref{reshalo}.

\begin{table}
\begin{center}
  \caption{Convergence Test Results for high-entropy case.  Listed are
    properties of the cold gas within $R_{200} = 206$ kpc after 7 Gyr
    of cooling. \label{table-conv}}
\begin{tabular}{||l|cccc|lll||}
\hline  \hline 
\, \, \, \, \, (1) & (2) & (3) & (4)\\
\, \, \, Name   & Radius &  M$_{cold}$  &  M$_{halo}$    \\
&      [kpc]    &  [$\Msun$] &  [$\Msun$]  \\
\hline 
high entropy low-resolution             &$ \sim R_{200}$      &       $ 2.9 \times 10^{10}$   &     $ 1.2 \times 10^{10}$  \\  
high entropy            &$ \sim R_{200}$      &       $ 3.0 \times 10^{10}$   &       $ 1.4 \times 10^{10}$ \\  
high entropy high-resolution             &$ \sim R_{200}$      &       $ 3.0 \times 10^{10}$  &  $ 1.4 \times 10^{10}$ \\

\hline 
\end{tabular}
\end{center}
{\small(2) Distance out to which cold particles can be found.  \\
  (3) Total cold mass (includes stars).   \\
  (4) Total cold gas mass in the halo, excluding the disc. }
\end{table}

\subsection{Evolution and dissolution of the clouds}

After $6$ Gyr of evolution of the high entropy high-resolution run, a variety of
cool clouds (temperature equal to the cut-off temperature in the
cooling function, $T_F=30 \,000$ K in most cases) are found, ranging from
compact clouds close to the disc (with masses 
$\sim 10^6 - 10^7$ and sizes of $r_{cloud}\sim0.7-1.3$ kpc) to diffuse 
clouds in the outer halo (typical masses of $\sim10^{7}$ M$_{\odot}$
and larger sizes of $r_{cloud}\sim4.5$ kpc).  
Clouds sizes are set by pressure confinement, therefore at fixed mass they are
 more diffuse at large radius, where the background hot gas pressure is lower.
 Because of the increased hot gas density towards the halo centre, we can resolve lower mass clouds more easily 
near the disc.

The fate of fragmentary clouds depends on various mechanisms (see also
Kaufmann et al. 2006, MB04).  One of the most important processes that influences
cloud survival is 
the Kelvin-Helmholtz (KH) instability. As a cool, dense cloud moves through
a hot, tenuous background, the interface between the two phases is 
subject to the growth of this instability, which can disrupt the clouds. At our resolution, 
the standard implementation of SPH does not resolve such instabilities due to
smoothing and the artificial viscosity which tends to blur any sharp
interface between the inner and outer medium (Agertz et al. 2007). For
the case in which self-gravity is unimportant, Murray et al. (1993)
derive a characteristic growth time for the instability of
\begin{equation}
\tau_{g}\approx\frac{r_{cloud}(\rho_{cloud}/\rho_{bg})^{0.5}}{U},
\end{equation} 
where $U$ is the relative velocity and $\rho_{bg}$ is the density of
the background medium. The clouds are therefore expected to break up
on time-scales comparable to $\tau_{g}$. In their numerical
experiments they actually found that the mass loss was still quite
small over time scales twice as long as $\tau_{g}.$ In our simulations
the density contrasts range from $\sim 10$ to $\sim 300$, the relative
velocities $U$ from $\sim 50$ to $\sim 160$ km/s and therefore we end
up with $\tau_{g}$ ranging from $\sim 0.3$ to $\sim 0.07$ Gyr. Based
on these estimates, the KH instability could dissolve a
significant fraction of the clouds with the parameters given above, if
they were evolved in isolation. But Vietri et al. (1997) showed that
with the inclusion of radiative cooling the effect of the
Kelvin-Helmholtz instability are stabilised and clouds in the mass
range given above survive. Moreover, Vieser \& Hensler (2007) showed that the
destructive effect of KH instability is significantly slowed if heat
conduction is included. 
We also note that the thermal instability can be
suppressed by buoyant oscillations in a medium with a pressure gradient
in thermal equilibrium (Balbus \& Soker 1989).  However our simulations (and real gaseous galactic haloes, for that matter)
are not in thermal equilibrium.   This suppression is also  reduced if convection occurs in the gas. 

Clearly, the issue of cloud survival remains an important one, especially for questions to do with the manner
of gas feeding onto galaxies.  The situation in simulations is more
dynamical than accounted for in most analytic explorations of KH: clouds grow by merging and new clouds are born due to the
thermal instability. Taken the Kelvin-Helmholtz instability together with
conductive evaporation, which can prevent the survival of small clouds
(see MB04), these effects would reduce the total mass in cool halo gas by
destroying the smallest clouds and modifying the outer layers of the
bigger clouds in our simulations\footnote{See also Kaufmann et al. (2006) for a discussion on cooling below $10^4$ K and self-shielding effects.}.   Interestingly, however, for the problem of over-cooling in
galaxy haloes, the disruption of cool clouds back into the hot halo can do nothing but help.

\bigskip 

In summary, the cool cloud mass spectrum that feeds disc formation 
is dependent on a variety of
physical processes: hydrodynamical instabilities, conduction
and the radiation field, none of which are currently resolved in these 
simulations (or any other cosmological simulations).  The initial cloud mass spectrum is determined by the initial 
perturbation spectrum, which is difficult to determine from first principles.
Encouragingly, however, we find that the total amount of gas
that cools into clouds seems to be independent of the initial perturbation spectrum.  
 The results we have presented 
on cloud properties can be seen as an upper limit on the total mass of cool clouds as 
most unresolved processes destroy clouds.  The details of how the clouds
would be observed is unfortunately more complicated because it depends 
on the temperature of the gas in the clouds and the ionizing background.  

\section{Summary and Conclusions}

We studied the cooling flow of gas within equilibrium NFW dark matter haloes, in order to mimic
the expected growth of $10^{12} \Msun$ halo galaxies after the cold-flow, rapid accretion period has ended for these systems ($z<1$).  
The two simulations contained the same baryonic mass and were identical other than their
initial gas density profiles -- one with a low central entropy that tracked the NFW density profile of the dark matter, 
and one with a high central entropy and low-density core in its density profile as might be expected in models
with substantial pre-heating feedback.
The simulations show dramatic
differences in their subsequent evolution. The high-entropy gas halo forms a fairly low-mass disc with high angular momentum
content, an extend halo cloud distribution, and an extended hot gas halo that produces an X-ray surface brightness consistent
with observations.  Most of the baryons in this case  reside in the
hot, quasi-stable gas halo. In contrast, the low entropy
run forms a smaller, but heavier disc quickly, shows no \hi \,in the
halo,  and has an X-ray luminosity that greatly exceeds observed limits for normal spiral galaxies.
These results have led us to three main conclusions:

\begin{itemize}

\item Changing the distribution of hot gas around galaxies, as might be expected in some pre-heating feedback schemes,
dramatically alters many aspects of the subsequent cooling behaviour, as well as the energy required to
stabilise the systems to subsequent cooling (see also Mo \& Mao 2002; Oh \& Benson 2003).  These differences are not only
manifest in the resultant optical characteristics of the central disc galaxy, but they may be tested
via X-ray observations and other non-optical probes of the haloes themselves.

\item Galaxy formation that proceeds via the thermal instability and cloud infall
 can dramatically affect  the properties of the resultant galaxy system.
  In our high entropy simulations, the  resultant disc is not only less massive, but it is larger, with more
specific angular momentum than in the low-entropy case.  Such behaviour
is simply not possible if the cooling occurs mainly near the disc in a central cooling flow.

\item It is possible to maintain a high-mass ($M \sim 5 \times 10^{10} \Msun$), 
low-density, quasi-stable gaseous
halo around a Milky Way size galaxy for a significant fraction of the Hubble
time without a continuous input of feedback energy.  Such an extended halo
(once in place) is a much better match to observations than 
low-entropy (cuspy/high-density) hot gas haloes that are commonly adopted
in first-order analytic explorations.   

\end{itemize}

These simulations show that the large core initial gas density profile
without any feedback can produce a galaxy that hardly suffers from many of the
classical problems in galaxy formation.  Kere{\v s} et al. (2009) have seen (smaller) cores in the density profiles
in some of their cosmological haloes and speculate that they could
be produced during the early chaotic, rapid assembly phase of haloes (without
the need of AGN heating or other preventive feedback). Another
possibility, explored by Tang et al. (2008) using one-dimensional simulations,
 is that AGN pre-heating is triggered by 
 the initial galaxy collapse itself.   This latter collapse could be associated with the end of the
 cold-mode phase (Kere{\v s} et al. 2009) or (perhaps equivalently) the end of the rapid accretion
 epoch (Wechsler et al. 2002).  As we have shown,
 once the hot gas density profile is suitably rearranged,
 it basically shuts down associated hot halo cooling for several Gigayears without
any further input of energy after that early phase.   Tests of these
ideas using high resolution numerical simulations in a full cosmological set-up are underway.

\section*{Acknowledgments}

The numerical simulations were performed on the IA64 Linux cluster at
the San Diego Supercomputer Center. We would like to thank Stelios
Kazantzidis for providing a code to generate isolated dark matter
haloes.  We acknowledge useful and stimulating discussions with Andrew
Benson, Hsiao-Wen Chen, Chris Churchill, Avishai Dekel, Jana Grcevich,
Dusan Kere{\v s}, Lucio Mayer, Julio Navarro, Peng Oh, Josh Peek, Mary
Putman, Jesper Rasmussen and Joop Schaye.  This work was supported by
the Center for Cosmology at UC Irvine. TK has been supported by the
Swiss National Science Foundation (SNF).

\end{document}